\documentclass[twocolumn,pra,psfig,showpacs,superscriptaddress]{revtex4}
\usepackage{amssymb}
\usepackage{amsfonts}
\usepackage{amsmath}
\usepackage{graphicx}

\begin{document}

\title{ H\'{e}non-Heiles Interaction for Hydrogen Atom in Phase Space}

\author{J.S. da Cruz Filho}
\email{josesilvacruz@uol.com.br}
\affiliation{Instituto de F\'{\i}sica, International Center of Physics, Universidade de
Bras\'{\i}lia, 70910-900, Bras\'\i lia, DF, Brazil}

\author{R.G.G. Amorim}
\affiliation{Faculdade Gama,Universidade de Bras\'{\i}lia, 72444-240, Bras\'{\i}lia, DF,
Brazil}
\affiliation{International Centre for Condensed Matter Physics Instituto de F\'{\i}sica,
Universidade de Bras\'{\i}lia, 70910-900, Bras\'{\i}lia, DF, Brazil}
\author{S.C. Ulhoa}
\affiliation{International Centre for Condensed Matter Physics Instituto de F\'{\i}sica,
Universidade de Bras\'{\i}lia, 70910-900, Bras\'{\i}lia, DF, Brazil}
\author{F.C. Khanna\footnote{Present address:Department o Physics and Astronomy,University of Victoria,Victoria B.C.Canada}}
\affiliation{Theoretical Physics Institute, University of Alberta, Edmonton, Alberta T6G
2J1, Canada}
\affiliation{TRIUMF, 4004, Westbrook Mall, Vancouver, British Columbia V6T 2A3, Canada}
\author{A. E. Santana}
\affiliation{Instituto de F\'{\i}sica, International Center of Physics, Universidade de
Bras\'{\i}lia, 70910-900, Bras\'\i lia, DF, Brazil}
\author{J.D.M. Vianna}
\affiliation{Instituto de F\'{\i}sica, Universidade Federal da Bahia, 40210-340,
Salvador, BA, Brazil}
\affiliation{Instituto de F\'{\i}sica, International Center of Physics, Universidade de
Bras\'{\i}lia, 70910-900, Bras\'\i lia, DF, Brazil}

\begin{abstract}
Using elements of symmetry, as gauge invariance, several aspects of a Schr%
\"{o}dinger equation represented in phase-space are introduced and analyzed
under physical basis. The Hydrogen atom is explored in the same context. Then we add a H\'enon-Heiles potential to the Hydrogen atom in order to explore chaotic features.

\end{abstract}

\maketitle

\section{Introduction}

In the sixties H\'enon and Heiles studied the third integral of motion in the context of a star orbiting a galaxy center\cite{HH}.
Until then it was known two of such constants of motion, angular momentum and energy. H\'enon-Heiles introduced a
combination of quadratic and cubic terms in the potential,
later it was found out that such a stellar dynamics could be chaotic or describe a regular orbit.
This close relation of H\'enon-Heiles potential to chaotic systems was also explored in quantum mechanics~\cite{HW,Ade}
and we intend to explore such features in phase-space.

It is interesting to describe representations of quantum equations directly in
phase-space; and some attempts have been made along these lines. As an example, Torres-Vega and Frederick \cite{vega1,vega2}, motivated by
the Husimi function, introduced a basis in the Hilbert space, $|\tau \rangle
=|q,p\rangle $, in phase space, $\Gamma $, such that the position and
momentum operators are given, respectively, by $\widehat{Q}=q/2+i\hbar
\partial /\partial p$ and $\widehat{P}=p/2-i\hbar \partial /\partial q$. The
Schr\"{o}dinger equation for bosons is then derived by taking the wave
function $|\psi (t)\rangle $ in $\Gamma $, i.e. $\psi (q,p;t)=\langle \tau
|\psi (t)\rangle $. This formalism has been applied, for example, to study
oscillators and to improve the harmonic analysis. Some physical aspects,
nevertheless, remain to be clarified. For instance, one problem is the
compatibility of the physical interpretation of the state $\psi (q,p;t)$, as
an amplitude of probability in $\Gamma $, using the Wigner function. Some of these difficulties have been solved by
using the notion of quasi-amplitude of probability, which is directly
associated with the Wigner function (the quasi-probability) and an analysis
of the symmetry group in phase space \cite{2kha1,2kha2}.

Representations of the Galilei group in a manifold with phase-space content
have been studied since long ago~\cite{schem1,schem2,schem3,schem4,schem5,schem6,schem7,schem8,schem9,schem10}.
Such representation, called symplectic unitary representation, has
been used by many groups and one interesting analysis is developed by using
the algebraic structure of the Wigner formalism~\cite{wig1,wig2,wig3,wig4,moy1,moy2}. In this approach, each operator, $A$,
defined in the Hilbert space, $\mathcal{H}$, is mapped in a function, $%
a_{W}(q,p)$, in $\Gamma $. The mapping $\Omega _{W}:A\rightarrow a_{W}(q,p)$%
, when applied to a product of operators is given by $\Omega
_{W}:AB\rightarrow a_{W}\star b_{W}$, where $\star $ is called the star (or
Moyal) product. The algebra of operators defined in $\mathcal{H}$ turns out
to be an associative (but not commutative) algebra in $\Gamma ,$ given by
the star product. This introduces a non-commutative algebraic structure in
phase space, a result that has been explored in different ways since the
paper by Wigner~\cite{2kha1,
2kha2,seb2,seb3,seb41,seb42,seb22,sig1,seb222,seb8,seb9,seb10,seb11,seb12,seb13,ron1,ron2,gosson1,gosson2,gosson3,gosson4}%
. A natural symplectic representation of Lie groups is introduced in $\Gamma
$ by considering star-operators defined as $\widehat{a}=a_{W}\ast $ \cite%
{2kha2,seb2}. For the Lorentz symmetry, the Klein-Gordon and Dirac equation
have been derived in $\Gamma $. These symplectic representations provide a
way to consider a perturbative approach for Wigner function on the bases of
symmetry groups. One example is the $\lambda \phi ^{4}$ field theory in
phase-space, leading to a relativisitic kinectic equation with a local
Boltzmann-like collision term. It is important to emphasize that, although
associated with the Wigner formalism, the symplectic representations have a
Hamiltonian, not a Liouvillian, operator as the generator of time
translations. This approach then provides satisfactory physical
interpretation for numerous aspects of a quantum theory formulated from a
unitary phase-space representation, in particular due to its clear
association with the Wigner function.

In the present work, the problem of constructing a formalism in phase space
based on unitary symmetry is addressed by using star-operators. Using the Galilei group, physical aspects of the
formalism are reviewed as the notion of quasi-amplitudes of probability in $%
\Gamma $. Then the Hydrogen atom  is described in pase space and the H\'enon-Heiles potential is added to such a system in order to explore chaotic features.

The paper is organized in the following way. In Section \ref{2} we present the
Schr\"{o}dinger in the phase space. In Section \ref{3}
we show the relation between the quasi-amplitude of probability and Wigner
function. In Sections \ref{4} and \ref{5}, we calculate the Wigner functions, as
applications, to the Hydrogen atom and we introduce a H\'enon-Heiles potential as an extra interaction in subatomic systems such as the Hydrogen atom  in a uniform magnetic field in phase space. In section \ref{nc} we introduce a parameter that can measure how the system is appart from the classical behavior.  Finally,
some closing comments are given in Section \ref{6}.

\section{Schr\"{o}dinger equation in phase space}\label{2}

Here we review some aspects of the Schr\"{o}dinger equation in phase space, in order to extend the
formalism to many-body systems. We consider initially a one-particle system
described by the Hamiltonian $H=\widehat{p}^{2}/2m$, where $m$ and $\widehat{%
p}$ are the mass and the momentum, respectively, of the particle. The Wigner
formalism for such a system is constructed from the Liouville-von Neumann
equation%
\begin{equation*}
i\hbar \partial _{t}\rho (t)=[H,\rho ],
\end{equation*}%
where $\rho (t)\;$\ is the density matrix. The Wigner function, $f_{w}(q,p),$
is defined by
\begin{equation}
f_{W}(q,p)=(2\pi \hbar )^{-3}\int dz\exp (\frac{ipz}{\hbar })\langle q-\frac{%
z}{2}|\rho |q+\frac{z}{2}\rangle ,  \label{wigago51}
\end{equation}%
and satisfies the equation of motion%
\begin{equation}
i\hbar \partial _{t}f_{W}(q,p,t)=\{H_{W},f_{W}\}_{M},  \label{wigago52}
\end{equation}%
where , $\{a,b\}_{M}=a\star b-b\star a$ is the Moyal bracket, such that the
star-product $a\star b$ is given by%
\begin{equation*}
a\star b=a(q,p)e^{\frac{i\hbar \Lambda }{2}}b(q,p)
\end{equation*}%
with $\Lambda =\overleftarrow{\partial }_{p}\overrightarrow{\partial }_{q}-%
\overleftarrow{\partial }_{q}\overrightarrow{\partial }_{p}.$ The functions $%
a(q,p)$ are defined in a manifold $\Gamma $, using the basis ($q,p$) with
the physical content of the phase space. In this formalism an operator, say $%
A,$ defined in the Hilbert space $\mathcal{H}$, is represented by the
function
\begin{equation*}
a_w(q,p)=\int dz\exp (\frac{ipz}{\hbar })\langle q-\frac{z}{2}|A|q+\frac{z}{2}%
\rangle ,
\end{equation*}%
such that the product of two operators, $AB$, reads%
\begin{equation*}
(AB)(q,p)=A(q,p)e^{\frac{i\hbar \Lambda }{2}}B(q,p)=A(q,p)\star B(q,p).
\end{equation*}%
The average of $A$ in a state $\psi \in \mathcal{H}$ is given by%
\begin{equation*}
\langle A\rangle =\langle \psi |A|\psi \rangle =\int
dqdpA(q,p)f_{W}(q,p)=Tr\rho A.
\end{equation*}

Due to the intricate structure of Eq. (\ref{wigago52}), one can look for an
alternative formulation for the Wigner function, in such a way that the usual
perturbative approach is extended to phase phase, in particular to describe
interacting many-body systems. Guided by this motivation, we proceed by
first introducing a Hilbert space, $\mathcal{H}(\Gamma )$, associated with
the phase space $\Gamma $. Consider the set of functions, $\psi (q,p)$ in $%
\Gamma $, such that $\int dpdq\phi ^{\ast }(q,p)\phi (q,p)<\infty $ is a
bilinear real form. Unitary mappings, $U(\alpha )$, in $\mathcal{H}(\Gamma )$
are naturally introduced by using the star-product, i.e. \ $U(\alpha )=\exp
(-i\alpha \widehat{A})$, where
\begin{align*}
\widehat{A}& =A(q,p)\star =A(q,p)\exp \left[ \frac{i\hbar }{2}(\frac{%
\overleftarrow{\partial }}{\partial q}\frac{\overrightarrow{\partial }}{%
\partial p}-\frac{\overleftarrow{\partial }}{\partial p}\frac{%
\overrightarrow{\partial }}{\partial q})\right]  \\
& =A(q+\frac{i\hbar }{2}\partial _{p},p-\frac{i\hbar }{2}\partial _{p}).
\end{align*}%
Let us consider some examples. For the basic functions in $\Gamma $, $q$ and
$p$ (3-dimensional Euclidian vectors), we have
\begin{equation}
\widehat{q}_{i}=q_{i}\star =q_{i}+\frac{i\hbar }{2}\partial _{p_{i}},
\label{eq 13}
\end{equation}%
\begin{equation}
\widehat{p}_{i}=p_{i}\star =p_{i}-\frac{i\hbar }{2}\partial _{q_{i}}.
\label{eq 14}
\end{equation}%
These operators satisfy the Heisenberg relations $\left[ \widehat{q}_{j},%
\widehat{p}_{l}\right] =i\hbar \delta _{jl}$. Then we introduce a Galilei
boost by defining the boost generator $\widehat{k}_{i}=mq_{i}\star
-tp_{i}\star =m\widehat{q}_{i}-t\widehat{p}_{i}$, $i=1,2,3$, such that
\begin{align*}
\exp \left( -i\mathbf{v}\cdot \widehat{\mathbf{k}}/\mathbb{\hbar }\right)
\widehat{q}_{j}\exp \left( i\mathbf{v}\cdot \widehat{\mathbf{k}}/\mathbb{%
\hbar }\right) & =\widehat{q}_{j}+v_{j}t\mathbf{,} \\
\exp \left( -i\mathbf{v}\cdot \widehat{\mathbf{k}}/\mathbb{\hbar }\right)
\widehat{p}_{j}\exp \left( i\mathbf{v}\cdot \widehat{\mathbf{k}}/\mathbb{%
\hbar }\right) & =\widehat{p}_{j}+mv_{j}\mathbf{.}
\end{align*}%
These results, with the commutation relations, show that $\widehat{q}$ and $%
\widehat{p}$ can be physically interpreted as the position and momentum
operators, respectively.

We introduce the operators $\overline{Q}\equiv(\overline{Q}_1,\overline{Q}_2,\overline{Q}_3)$ and $\overline{P}\equiv (\overline{P}_1, \overline{P}_2, \overline{P}_3)$, such that $[%
\overline{Q}_i,\overline{P}_j]=0$, $\ \overline{Q}_i|q,p\rangle=q_i|q,p\rangle$ and $%
\overline{P}_i|q,p\rangle=p_i|q,p\rangle$, with
\begin{equation}
\langle q,p|q^{\prime},p^{\prime}\rangle=\delta(q-q^{\prime})\delta
(p-p^{\prime}),  \notag
\end{equation}
and $\int dqdp|q,p\rangle\langle q,p|=1$. From a physical point of view, we
observe the transformation rules:
\begin{equation*}
\exp(-iv \cdot\frac{\widehat{K}}{\hbar})2\overline{Q}_i\exp(iv\cdot\frac{\widehat{K}}{%
\hbar})=2\overline{Q}_i+v_it\mathbf{1},
\end{equation*}
and
\begin{equation*}
\exp(-iv\cdot\frac{\widehat{K}}{\hbar})2\overline{P}_i\exp(iv\cdot\frac{\widehat{K}}{%
\hbar})=2\overline{P}_i+mv_i\mathbf{1}.
\end{equation*}
Then $\overline{Q}$ and $\overline{P}$ are transformed, under the Galilei
boost, as position and momentum, respectively. Therefore, the manifold
defined by the set of eingenvalues $\{q,p\}$ has the content of a phase
space. However, the operators $\overline{Q}$ and $\overline{P}$ are not
observables, since they commute with each other.

Considering a homogeneous systems satisfying the Galilei symmetry, the
commutations relation between $\widehat{K}$ and $\widehat{H}$ is $[\widehat{K%
}_{j},\widehat{H}]=i\widehat{P}_{j}$, i.e.%
\begin{equation*}
\lbrack mq_{j}+i\frac{\partial }{\partial p_{j}},H(q,p)\star ]=ip_{j}+\frac{\hbar%
}{2}\frac{\partial }{\partial q_{j}}.
\end{equation*}%
A solution, providing a general form to $\widehat{H}=H(q,p)\star $, is
\begin{align}
\widehat{H}& =\frac{p^{2}\star }{2m}+V(q)\star   \notag \\
& ={\frac{p^{2}}{2m}}-{\frac{\hbar ^{2}}{8m}}{\frac{\partial ^{2}}{\partial
q^{2}}}-{\frac{i\hbar p}{2m}}{\frac{\partial }{\partial q}}+V(q\star ){.}
\label{ago20.2}
\end{align}%
This is the Hamiltonian of a one-body system in an external field. Such an
interpretations has to be consistent with the notion of gauge invariance.
Let us investigate this point by analysing the Lagrangian associated with
the equation of motion for $\psi (q,p;t)=\langle q,p|\psi (t)\rangle .$

Consider the time evolution of a state $\psi (q,p;t)$, that is given by $%
\psi (q,p;t)=U(t,t_{0})\psi (q,p;t_{0}),$ where $U(t,t_{0})=\exp (-i\hbar
(t-t_{0})\widehat{H})$. This result leads to a Schr\"{o}dinger-like equation
in phase-space, i.e.%
\begin{equation}
i\hbar \partial _{t}\psi (q,p,t)=\widehat{H}\psi (q,p,t)  \label{ago25.1}
\end{equation}%

\section{Quasi-amplitude of probability and Wigner function}\label{3}

Now let us consider the physical meaning of the state $\psi (q,p,t)$. This
is carried out by associating $\psi (q,p,t)$ with the Wigner function. From
Eq. (\ref{ago25.1}), one can prove that $f(q,p,t)=\psi (q,p,t)\star \psi
^{\dagger }(q,p,t)$ satisfies Eq. (\ref{wigago52}) \cite{2kha2,seb2}. In
addition, using the associative property of the Moyal product and the
relation
\begin{equation*}
\int dqdp\psi (q,p,t)\star \psi ^{\dagger }(q,p,t)=\int dqdp\psi (q,p,t)\psi
^{\dagger }(q,p,t),
\end{equation*}%
we have
\begin{align*}
\langle A\rangle & =\langle \psi |A|\psi \rangle  \\
& =\int dqdp\psi (q,p,t)\widehat{A}(q,p)\psi ^{\dagger }(q,p,t) \\
& =\int dqdpf_{W}(q,p,t)A(q,p,t),
\end{align*}%
where $\widehat{A}(q,p)=A(q,p)\star $ \ is an observable. Thus the Wigner
function is calculated by using
\begin{equation}
f_{W}(q,p)=\psi (q,p)\star \psi ^{\dagger }(q,p).  \label{wig1}
\end{equation}%
It is to be noted also that the eigenvalue equation,
\begin{equation}
H(q,p)\star \psi =E\psi ,
\end{equation}%
results in $H(q,p)\star f_{W}=Ef_{W}.$ Therefore, $\psi (q,p)$ and $%
f_{W}(q,p)$ satisfy the same differential equation. These results show that
Eq. (\ref{ago25.1}) is a fundamental starting point for the description of
quantum physics in phase space, fully compatible with the Wigner formalism.
An attractive aspect of this method is that the powerfull methods developed
in quantum theory based on the notion of linear spaces, such as the
perturbative techniques, can be explored in the phase space in a
straightforward way. The fundamental tool is the phase-space wave function $%
\psi (q,p,t)$, that is a quasi-amplitude of probability.

\section{Hydrogen atom in phase space}\label{4}

In this section we solve the Schr\"{o}dinger equation in phase space for the
Coulomb potential. Our analysis considers a one dimensional system, where
the potential is $V(q)\star =V(\widehat{q})=\frac{-Ze^{2}}{4\pi \epsilon
_{0}|q|}\star $ and leading to
\begin{equation}
(\frac{p^{2}}{2m}-\frac{-Ze^{2}}{4\pi \epsilon _{0}|q|})\star \psi
(q,p)=E\psi (q,p).  \label{jun20131}
\end{equation}%
We follow in parallel to previous developments \cite{hydo1, hydo2, hydo3}.
However, we have to note that the difference in our operators is a necessary
conditions to provide the physical interpretation for the wave function $%
\psi (q,p)$, a quasi-amplitude of probability, leading to the Wigner
function. This is the main (and new) result to be derived here. Equation (%
\ref{jun20131}) is solved in two distinct regions, $q>0$ and $q<0$. We
denote $\psi _{>}(q,p)$ the solution for $q>0$ and $\psi _{<}(q,p)$ the
solution for $q<0$. We calculate for the case $\psi _{>}(q,p)$ and assume
that a similar calculation for $\psi _{<}(q,p)$ can be carried out. The
equation for $\psi _{>}(q,p)$ is
\begin{equation}
\frac{1}{2m}(p-\frac{i\hbar }{2}\partial _{q})^{2}\psi _{>}(q,p)-\frac{Ze^{2}%
}{4\pi \epsilon _{0}}(q+\frac{i\hbar }{2}\partial _{p})^{-1}\psi
_{>}(q,p)=E\psi _{>}(q,p).  \label{+eq}
\end{equation}%
Assuming that $\psi _{>}(q,p)=\exp {(\frac{-2iqp}{\hbar })}\phi (q,p)$, and
using the relations
\begin{equation}
\exp {(\frac{-2iqp}{\hbar })}(\frac{-i\hbar }{2}\partial _{q})\exp {(\frac{%
2iqp}{\hbar })}=(p-\frac{i\hbar }{2}\partial _{q}),  \notag
\end{equation}%
\begin{equation}
\exp {(\frac{-2iqp}{\hbar })}(\frac{-i\hbar }{2}\partial _{q})^{2}\exp {(%
\frac{2iqp}{\hbar })}=(p-\frac{i\hbar }{2}\partial _{q})^{2},  \notag
\end{equation}%
and
\begin{equation}
\exp {(\frac{-2iqp}{\hbar })}A(2q+\frac{i\hbar }{2}\partial _{p})\exp {(%
\frac{2iqp}{\hbar })}=A(q+\frac{i\hbar }{2}\partial _{p}),  \notag
\end{equation}%
Eq.(\ref{+eq}) becomes
\begin{equation}
\lbrack \frac{-\hbar ^{2}}{8m}\partial _{q}^{2}-\frac{Ze^{2}}{4\pi \epsilon
_{0}}(2q+\frac{i\hbar }{2}\partial _{p})^{-1}]\phi (q,p)=E\phi (q,p).
\label{hys1}
\end{equation}

Changing variables to $\eta =2q+\frac{i\hbar }{2}\partial _{p}$, we have $%
\partial _{q}^{2}\phi =4\partial _{\eta }^{2}\phi $, where $\phi =\phi (\eta
,p)$. Then Eq. (\ref{hys1}) takes the form,
\begin{equation}
\lbrack \frac{-\hbar ^{2}}{2m}\partial _{\eta }^{2}-\frac{Ze^{2}}{4\pi
\epsilon _{0}}\frac{1}{\eta }]\phi (\eta ,p)=E\phi (\eta ,p).  \label{hys2}
\end{equation}%
Using the ansatz \cite{hydo1} $\phi (\eta ,p)=\eta \exp {(\frac{-\eta }{2})}%
\omega (\eta ,p)$ we get
\begin{equation}
\eta \partial _{\eta }^{2}\omega +(2-\eta )\partial _{\eta }\omega -(1-\frac{%
2mZe^{2}}{4\pi \epsilon _{0}\hbar ^{2}})\omega =0,  \label{hys3}
\end{equation}%
where $E=\frac{-\hbar ^{2}}{8m}$. Defining $\gamma =\frac{2\hbar ^{2}\pi
\epsilon _{0}}{mZe^{2}}$, we have

\begin{equation}
\eta \partial _{\eta }^{2}\omega +(2-\eta )\partial _{\eta }\omega -(1-\frac{%
1}{\gamma })\omega =0.  \notag
\end{equation}%
This differential equation is identified with the equation for confluent
hypergeometric functions. Its solution in the variable $\eta $ is given by
\cite{hydo1}
\begin{equation}
\omega (\eta ,p)=F(1-n,2;\eta )h(p),
\end{equation}%
where $F(1-n,2;\eta )$ is the confluent hypergeometric function and $h(p)$
is an arbitrary function of variable $p$.

We using $n=\frac{1}{\gamma }$, the energy has the form
\begin{equation}
E=-\frac{mZe^{4}}{(4\pi \epsilon _{0})^{2}2\hbar ^{2}}\frac{1}{n^{2}}.
\end{equation}%
which can be put into the form $E=-13,6\frac{1}{n^{2}}eV$, as expected.

Finally we write Eq.(\ref{+eq}) as,

\begin{equation}
\psi _{n_{>}}(\eta ,p)=h(p)\exp {(\frac{-2iqp}{\hbar })}\eta \exp {(\frac{%
-\eta }{2})}F(1-n,2;\eta ).
\end{equation}%
And analogously, for $\psi _{<}(q,p)$ the solution is
\begin{equation}
\psi _{n_{<}}(\eta ,p)=h(p)\exp {(\frac{-2iqp}{\hbar })}(-\eta )\exp {(\frac{%
-\eta }{2})}F(1-n,2;-\eta ).
\end{equation}%
For the particular case of $h(p)=\frac{1}{\sqrt{2\pi }}\exp {(\frac{8i\pi
\epsilon _{0}\hbar }{me^{2}})p}$, and using the relation
\begin{equation}
(q+\frac{i\hbar }{2}\partial _{p})^{n}h(p)=(q-\frac{4\pi \epsilon _{0}\hbar
^{2}}{me^{2}})^{n}h(p),
\end{equation}%
the solutions are

\begin{align*}
\psi _{n_{>}}(q,p)& =\exp {(\frac{-2iqp}{\hbar })}(q-\frac{4\pi \epsilon
_{0}\hbar ^{2}}{me^{2}}) \\
& \times \exp {(\frac{-1}{n}(q-\frac{4\pi \epsilon _{0}\hbar ^{2}}{me^{2}})}%
)F(1-n,2;\frac{2(q-\frac{4\pi \epsilon _{0}\hbar ^{2}}{me^{2}})}{n}) \\
& \times \frac{1}{\sqrt{2\pi }}\exp {(\frac{8i\pi \epsilon _{0}\hbar ^{2}}{%
me^{2}}p}).
\end{align*}%
This solution is similar to the results of Ref. \cite{hydo3}. But in our
construction, we get the physical interpretation in terms of the Wigner
function. For example, taking $n=1$, the Wigner function for the fundamental
state is
\begin{equation}
f_{W}{}_{1_{>}}(q,p)=\psi _{1_{>}}(q,p)\star \psi _{1_{>}}^{\ast }(q,p).
\end{equation}%
Computing the star product to second order we get

\begin{equation}
f_{W}{}_{1_{>}}(q,p)\simeq(q-\frac{4\pi\epsilon_{0}\hbar^{2}}{me^{2}}%
)^{2}\exp{-2(q-\frac{4\pi\epsilon_{0}\hbar^{2}}{me^{2}})}.
\end{equation}

Using this Wigner function, we find the maximum of the probability density
associated with the position variable. It is interesting to take the
expression of the probability density, as

\begin{equation}
\sigma(q)=\int dp\psi(q,p)\star\psi^{\ast}(q,p).
\end{equation}
Next by differentiating with respect to $q$, we get $q=4\pi\epsilon_{0}\hbar
^{2}/me^{2},$ which is the Bohr radius.

\section{H\'enon-Heiles interaction for hydrogen atom in phase space}\label{5}
In this section we'll allow a new interaction in the Hamiltonian of the Hydrogen atom in the presence of magnetic field which is realized by H\'enon-Heiles potential. In general subatomic systems
such as the Hydrogen atom can be described by a harmonic oscillator through a proper change of variables, however this procedure is limited.
Let us consider a Hamiltonian of the form~\cite{Friedrich,Ade}

\begin{equation}
\mathcal{H}= \frac{1}{2}\left(p^{2}_{x}+ p^{2}_{y}\right)+
\lambda(V_1+V_2) \label{eq:1.1},
\end{equation}%

where
\begin{equation}
V_1= \frac{1}{2}\left( q_{x}^{2}+q_{y}^{2} \right) + q_{x}^{2}q_{y} - \frac{1}{3}q_{y}^{3},  \notag
\end{equation}%
and
\begin{equation}
V_2=\frac{5}{2}q_{x}^{2}q_{y}^{2}\left(q_{x}^{2}+ q_{y}^{2}\right) -\varepsilon \left(q_{x}^{2}+ q_{y}^{2}\right).  \notag
\end{equation}%
This Hamiltonian describes the system we are interested in where $V_1$ is the H\'enon-Heiles potential and $V_2$ stands for the potential of
the Hydrogen atom in an uniform magnetic field potential. The term $ \varepsilon=E\gamma^{-\frac{2}{3}}$ is scaled
energy and measures the energy of the electron in units of magnetic field
which governs the classical dynamics. The quantity $ \gamma=\hbar\omega/\Re$ is  the magnetic field in atomic units of $2.35 \times 10^5 T$
where $\Re$ is the Rydeberg energy.

In order to represent this system in phase space we perform in (\ref{eq:1.1}) a change of variables $(q\rightarrow\hat{q}; p\rightarrow\hat{p})$ wich are given by
the operators defined in (\ref{eq 13}) and (\ref{eq 14}). Thus we define the annihilation and creation operators in each direction with $\hbar=1$ which read

\begin{equation}
\widehat{A}=\frac{1}{\sqrt{2}} \left(\widehat{q}_{x}+i\widehat{p}_{x}\right)  \quad\mbox{;}\quad \widehat{A}^\dag=\frac{1}{\sqrt{2}}\left(\widehat{q}_{x}-i\widehat{p}_{x}\right) \notag
\end{equation}
and
\begin{equation}
\widehat{B}=\frac{1}{\sqrt{2}}\left(\widehat{q}_{y}+i\widehat{p}_{y}\right)  \quad\mbox{;}\quad \widehat{B}^\dag=\frac{1}{\sqrt{2}}\left(\widehat{q}_{y}-i\widehat{p}_{y}\right)\,. \notag
\end{equation}
Hence the Hamiltonian is given by
\begin{widetext}
\begin{eqnarray}
\widehat{H}(q,p)&=& \left(\widehat{A}\widehat{A}^\dag-\frac{1}{2}\right)+
\left(\widehat{B}\widehat{B}^\dag-\frac{1}{2}\right)+ \lambda\Big[\frac{1}{2\sqrt{2}}\left(\widehat{A}+\widehat{A}^\dag\right)^{2}\left(\widehat{B}+
\widehat{B}^\dag\right)-\frac{1}{6\sqrt{2}}\left(\widehat{B}+\widehat{B}^\dag\right)^{3}\nonumber \\
&+&\frac{5}{2} \left( \frac{1}{4} \left(\widehat{A}+\widehat{A}^\dag\right)^{2}
  \left(\widehat{B}+\widehat{B}^\dag\right)^{2} \left(\frac{1}{2} \left( \widehat{A}+\widehat{A}^\dag\right)^{2} + \frac{1}{2}\left( \widehat{B}+\widehat{B}^\dag\right)^{2} \right) \right) - \varepsilon \left( \frac{1}{2} \left( \widehat{A}+\widehat{A}^\dag\right)^{2} + \frac{1}{2}\left( \widehat{B}+\widehat{B}^\dag\right)^{2} \right)\Big]\nonumber\\
 &=& \widehat{H}^{0}(q,p) +  \widehat{V}
\end{eqnarray}
\end{widetext}
where

\begin{equation}
\widehat{H}^{0}(q,p)= \left(\widehat{A}\widehat{A}^\dag-\frac{1}{2}\right)+
\left(\widehat{B}\widehat{B}^\dag-\frac{1}{2}\right) \label{eq:1.2a}
\end{equation}
and
\begin{small}
\begin{eqnarray}
\widehat{V}&=&  \frac{1}{2\sqrt{2}}\left(\widehat{A}+\widehat{A}^\dag\right)^{2}\left(\widehat{B}+
\widehat{B}^\dag\right)-\frac{1}{6\sqrt{2}}\left(\widehat{B}+\widehat{B}^\dag\right)^{3}\nonumber \\
&+&\frac{5}{16} \left(\widehat{A}+\widehat{A}^\dag\right)^{2}
  \left(\widehat{B}+\widehat{B}^\dag\right)^{2} \left( \left( \widehat{A}+\widehat{A}^\dag\right)^{2} + \left( \widehat{B}+\widehat{B}^\dag\right)^{2} \right) \nonumber \\
 &-&\varepsilon\left( \frac{1}{2} \left( \widehat{A}+\widehat{A}^\dag\right)^{2} + \frac{1}{2}\left( \widehat{B}+\widehat{B}^\dag\right)^{2} \right). \label{eq:1.2b}
\end{eqnarray}
\end{small}
It is interesting to note that such a Hamiltonian describes a system of coupled oscillators.

We'll use the perturbation theory to treat this system in order to calculate quasi-probability amplitudes and respective Wigner functions. Thus we'll restrict our attention to the case where $\widehat{V}$ is a perturbation. Let us begin with the time-independent Schr\"odinger equation in phase space for harmonic
oscillator which reads

\begin{equation}
\widehat{\mathcal{H}}(q,p)^{0}\psi_{n}^{0}= E_n^{0}\psi_{n}^{0},\notag
\end{equation}%
where
\begin{eqnarray}
\psi_{n}^{0}(q,p)&=& \psi_{nx}^{0}(q,p)\psi_{ny}^{0}(q,p)\notag\\
&=&\sqrt{\frac{e}{\pi}} \frac{2^n}{ \sqrt{n!}} (a^{\dag})^{n} exp\left( \frac{-2h(q,p)}{ \omega}   \right).\notag
\end{eqnarray}%
and
\begin{equation}
 E_n^{0}=(n_x +1/2)\omega + (n_y +1/2)\omega,\notag
\end{equation}%

We recall that the Hamiltonian is
\begin{equation}
\widehat{H}(q,p)=\widehat{H}^{0}(q,p) + \widehat{V} \label{eq:1.3}
\end{equation}
where $\widehat{H}^{0}(q,p) $ is the unperturbed
Hamiltonian and  $ \widehat{V}$ is the perturbation.
Therefore the wave function and energy have a perturbative series as

\begin{equation}
\psi_{n}=\psi_{n}^{0}+  \psi_{n}^{1}+  \psi_{n}^{2}+\cdots \label{eq:1.3a}
\end{equation}%
and
\begin{equation}
E_{n}=E^{0}+  E_{n}^{1}+  E_{n}^{2}+\cdots \,.\label{eq:1.3b}
\end{equation}
If we insert equations (\ref{eq:1.3a}) and (\ref{eq:1.3b}) into equation (\ref{eq:1.3}), then
it yields
\begin{eqnarray}
\widehat{H}(q,p)\psi_{n}&=&\left(\widehat{H}^{0}(q,p) + \widehat{V}\right) \left[ \psi_{n}^{0}+  \psi_{n}^{1}+ \psi_{n}^{2}+\cdots \right] \nonumber\\
&=&\left(E^{0}+  E_{n}^{1}+  E_{n}^{2}+\cdots\right) \big[ \psi_{n}^{0}+  \psi_{n}^{1}\nonumber\\
&+&  \psi_{n}^{2}+\cdots \big] \,.\nonumber
\end{eqnarray}
Thus the zero-order term is

\begin{equation}
\widehat{H}^{0}(q,p)\psi_{n}^{0}= E_{n}^{0}\psi_{n}^{0} \,,\label{eq:1.4}
\end{equation}
the first order correction  is
\begin{equation}
\widehat{H}^{0}(q,p)\psi_{n}^{1}+ \widehat{V}\psi_{n}^{0}
= E_{n}^{0}\psi_{n}^{1} + E_{n}^{1}\psi_{n}^{0} \label{eq:1.4a}
\end{equation}
and the second order correction is given by

\begin{equation}
\widehat{H}^{(0)}(q,p)\psi_{n}^{2}+\widehat{V}\psi_{n}^{1}
= E_{n}^{0}\psi_{n}^{2}+E_{n}^{1}\psi_{n}^{1}+ E_{n}^{2}\psi_{n}^{0} \label{eq:1.4b}
\end{equation}

Now we intend to develop separated these last two corrections in the following subsections.

\subsection{First order correction}
The first order corrections are obtained by writing Eq.(\ref{eq:1.4a}) as
\begin{equation}
(\widehat{H}^{0}(q,p)-E_{n}^{0})\psi_{n}^{1}= (E_{n}^{1}- \widehat{V})\psi_{n}^{0}, \label{eq:1.5}
\end{equation}
then $\psi_{n}^{1}$ is expressed in terms of of a complete set of states as
\begin{equation}
\psi_{n}^{1}= \sum_{m\neq n}a_{m}^{1}\psi_{m}^{0} \,.\label{eq:1.5a}
\end{equation}
If we substitute (\ref{eq:1.5a}) into (\ref{eq:1.5}), then we get

\begin{equation*}
\sum_{m\neq n}(E_{m}^{0}-E_{n}^{0}) a_{m}^{1}\psi_{m}^{0}
= (E_{n}^{1}- \widehat{V})\psi_{n}^{0}\,.
\end{equation*}
Multiplying by $\psi_{k}^{\dag 0}(q,p)$ and integrating in $dpdq$, yields

\begin{eqnarray}
&&\sum_{m\neq n}  a_{m}^{1} (E_{m}^{0}-E_{n}^{0})\int \psi_{k}^{\dag 0} \star \psi_{m}^{0} dpdq\nonumber\\
&=& \int \psi_{k}^{\dag 0}  (E_{n}^{1}- \widehat{V})\psi_{n}^{0}dpdq\,.  \label{eq:1.6}
\end{eqnarray}
We recall that the wave functions of unperturbed systems are ortogonal i.e

\begin{equation}
\int \psi_{k}^{\dag 0} \star \psi_{m}^{0} dpdq
= \int \psi_{k}^{\dag 0} \psi_{n}^{0}dpdq=\delta_{km},  \nonumber
\end{equation}
then equation (\ref{eq:1.6}) becomes

\begin{equation*}
\sum_{k\neq n}  a_{k}^{1} (E_{k}^{0}-E_{n}^{0})
= E_{n}^{1}\delta_{kn} - \int \psi_{k}^{\dag 0} \widehat{V} \psi_{n}^{0}dpdq\,.
\end{equation*}
It is worth noting that two cases arise.  $k=n$ yields
\begin{equation}
E_{n}^{1} = \int \psi_{n}^{\dag 0} \widehat{V} \psi_{n}^{0}dpdq\,, \label{eq:1.61}
\end{equation}
such an equation provides the first order correction
to the energy of the unperturbed system. On the other hand the second condition $k\neq n$ gives us
\begin{equation}
\psi_{n}^{1}= \sum_{m\neq n} \int \frac{\psi_{m}^{\dag 0} \widehat{V} \psi_{n}^{0}}{ E_{n}^{0} - E_{m}^{0} }dpdq \psi_{m}^{0}\,. \label{eq:1.7}
\end{equation}
The above expression is well defined since $m\neq n$. Therefore the first order approximation
of the wave function for the  H\'enon-Heiles and hydrogen atom potential (\ref{eq:1.2b}) is $\psi_{n}=\psi_{n}^{(0)}+\psi_{n}^{1}$
where $\psi_{n}^{1}\equiv \psi_{n_{x}}^{1}(q_{x},p_{x})\psi_{n_{y}}^{1}(q_{y},p_{y})$. It reads
\begin{small}
\begin{equation}
\psi_{n}^{1}= \frac{1}{4%
\sqrt{2}} a0 -\frac{1}{6\sqrt{2}} a1 + \frac{5}{16}(a2+a3+a4+a5+a6) -\frac{\varepsilon}{2} a7 \label{eq:1.7c}
\end{equation}
\end{small}
where the coefficients are listed in appendix \ref{appendix}.

\begin{figure}[htb!]
\includegraphics[scale=0.3]{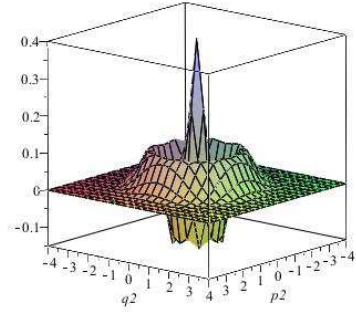}
\caption{Winger function n=0}
\label{fig1}
\end{figure}

\begin{figure}[htb!]
\includegraphics[scale=0.3]{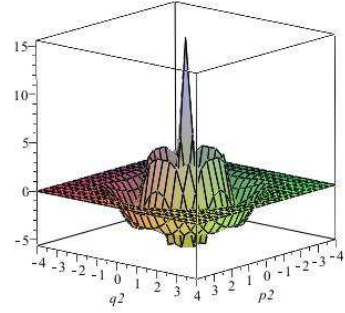}
\caption{Winger function n=2}
\label{fig2}
\end{figure}

\begin{figure}[htb!]
\includegraphics[scale=0.3]{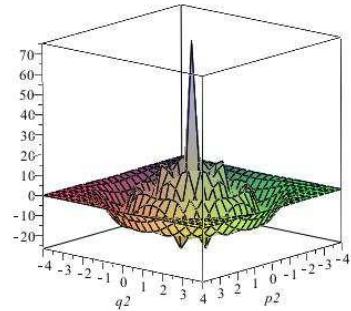}
\caption{Winger function n=4}
\label{fig3}
\end{figure}

\begin{figure}[htb!]
\includegraphics[scale=0.3]{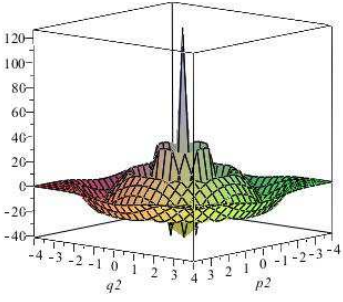}
\caption{Winger function n=6}
\label{fig4}
\end{figure}

\begin{figure}[htb!]
\includegraphics[scale=0.3]{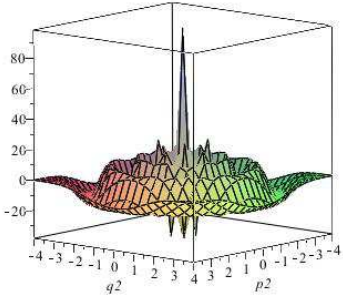}
\caption{Winger function n=8}
\label{fig5}
\end{figure}

\begin{figure}[htb!]
\includegraphics[scale=0.3]{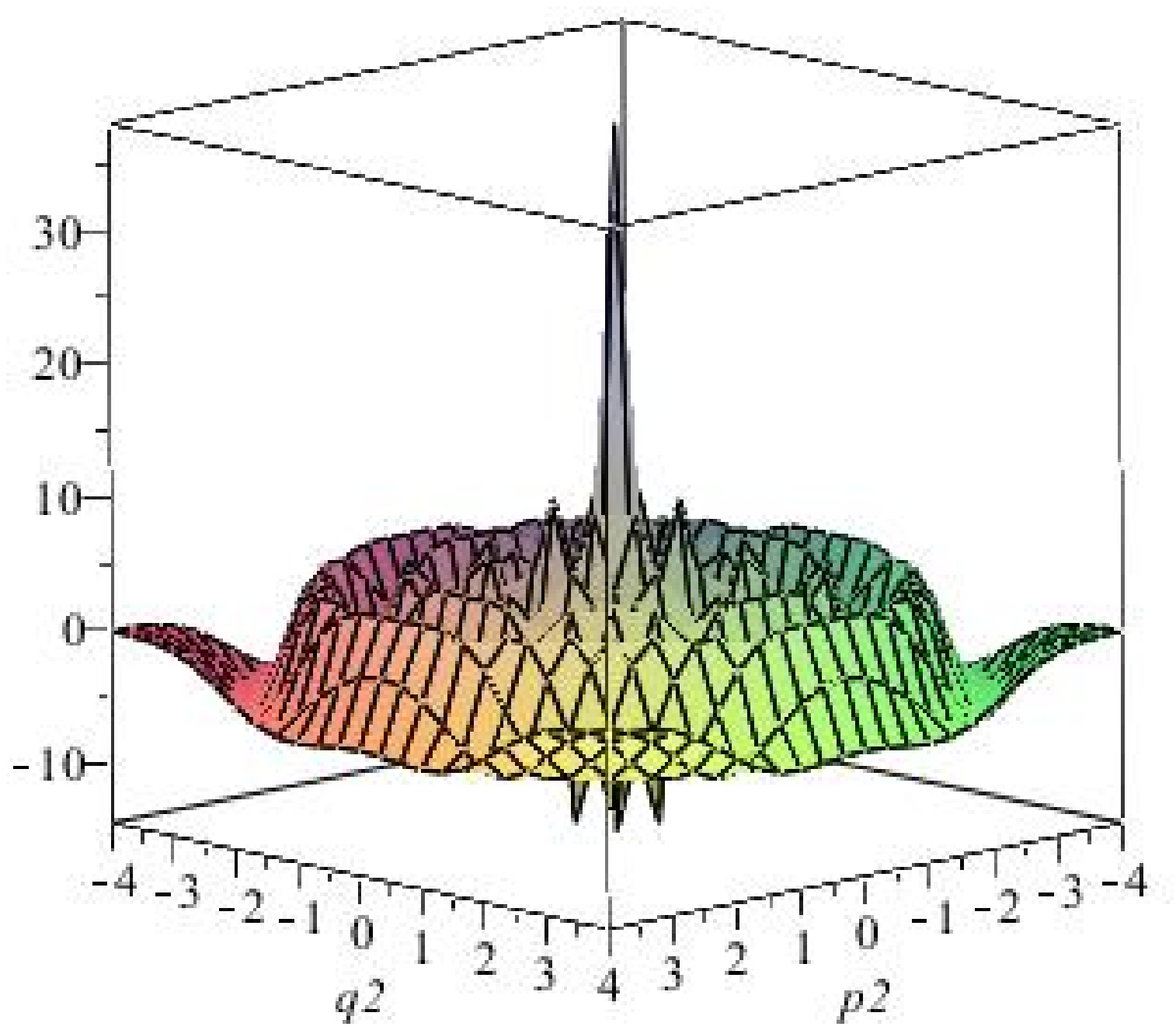}
\caption{Winger function n=10}
\label{fig6}
\end{figure}

Finaly the Wigner function is given by
\begin{small}
\begin{equation*}
f_{n}^{1}(q_{x},p_{x},q_{y},p_{y})=\psi _{n_{x}}(q_{x},p_{x})\psi
_{n_{y}}(q_{y},p_{y})\ast \psi _{n_{x}}^{\dagger }(q_{x},p_{x})\psi
_{n_{y}}^{\dagger }(q_{y},p_{y}).
\end{equation*}%
\end{small}
We present some numerical results in figures FIG.\ref{fig1}, FIG.\ref{fig2}, FIG.\ref{fig3}, FIG.\ref{fig4}, FIG.\ref{fig5} and FIG.\ref{fig6}. We also present such results on tables
TABLE \ref{tbl:tablelabel1}, TABLE \ref{tbl:tablelabel2}, TABLE \ref{tbl:tablelabel3}, TABLE \ref{tbl:tablelabel4}, TABLE \ref{tbl:tablelabel5} and TABLE \ref{tbl:tablelabel6}.



\begin{table}
\caption{Winger function n=0.}
\centering
\begin{tabular}{lll}
\hline
$\varepsilon$ & Maximum & Minimum  \\
\hline
0 & 0.082 & -0.023\\
0.28 & 0.1 & -0.03\\
0.5 & 0.162 & -0.055\\
1 & 0.4 & -0.151\\
\hline
\end{tabular}
\label{tbl:tablelabel1}
\end{table}

\begin{table}
\caption{Winger function n=2.}
\centering
\begin{tabular}{lll}
\hline
$\varepsilon$ & Maximum & Minimum  \\
\hline
0 & 12.984 & -4.803\\
0.28 & 13.187 & -4.869\\
0.5 & 13.63 & -5.013\\
1 & 15.57 & -5.644\\
\hline
\end{tabular}
\label{tbl:tablelabel2}
\end{table}

\begin{table}
\caption{Winger function n=4.}
\centering
\begin{tabular}{lll}
\hline
$\varepsilon$ & Maximum & Minimum  \\
\hline
0 & 69.986 & -25.131 \\
0.28 & 70.402 & -25.233 \\
0.5 & 71.313 & -25.458 \\
1 & 75.291 & -26.438\\
\hline
\end{tabular}
\label{tbl:tablelabel3}
\end{table}

\begin{table}
\caption{Winger function n=6.}
\centering
\begin{tabular}{lll}
\hline
$\varepsilon$ & Maximum & Minimum  \\
\hline
0 & 118.512 & -39.587 \\
0.28 & 119.153 & -39.741 \\
0.5 & 120.556 & -40.078 \\
1 & 126.689 & -41.551  \\
\hline
\end{tabular}
\label{tbl:tablelabel4}
\end{table}

\begin{table}
\caption{Winger function n=8.}
\centering
\begin{tabular}{lll}
\hline
$\varepsilon$ & Maximum & Minimum  \\
\hline
0 & 91.654 & -35.752 \\
0.28 & 92.192 &	-35.948 \\
0.5 & 93.368 & -36.375 \\
1 & 98.51 &	-38.244\\
\hline
\end{tabular}
\label{tbl:tablelabel5}
\end{table}

\begin{table}
\caption{Winger function n=10.}
\centering
\begin{tabular}{lll}
\hline
$\varepsilon$ & Maximum & Minimum  \\
\hline
0 & 35 & -13.229  \\
0.28 & 35.235 &	-13.323 \\
0.5 & 35.751 & -13.529 \\
1 & 38.002 & -14.427\\
\hline
\end{tabular}
\label{tbl:tablelabel6}
\end{table}

\subsection{Second order correction}
To establish second order corrections we make use of Eq.((\ref{eq:1.4b})). In order to obtain these corrections in terms of
\begin{equation}
(\widehat{H}^{0}(q,p)-E_{n}^{0})\psi_{n}^{2} = -(\widehat{V}-E_{n}^{1})\psi_{n}^{1}+ E_{n}^{2}\psi_{n}^{0}\,, \label{eq:1.8}
\end{equation}
 which means
\begin{equation}
\psi_{n}^{1}= \sum_{m\neq n}a_{m}^{1}\psi_{m}^{0} \label{eq:1.8a}
\end{equation}
and
\begin{equation}
\psi_{n}^{2}= \sum_{m\neq n}a_{m}^{2}\psi_{m}^{0}\,. \label{eq:1.8b}
\end{equation}
If we use such dependencies into equation (\ref{eq:1.8}) then we have
\begin{eqnarray}
\sum_{m\neq n} (E_{m}^{0}-E_{n}^{0})a_{m}^{2}\psi_{m}^{0} &=&
 -\sum_{m\neq n}\Big[(\widehat{V}-E_{n}^{1})a_{m}^{1}\psi_{m}^{0} \nonumber\\
 &+& E_{n}^{2}\psi_{n}^{0}\Big]\,, \notag
\end{eqnarray}
now we multiply the entire equation by $\psi_{k}^{\dag\, 0}(q,p)$ and integrate in $dpdq$, thus it yields
\begin{eqnarray}
\sum_{k\neq n}a_{k}^{2} (E_{k}^{0}-E_{n}^{0})&=&
 -\sum_{m\neq n}a_{m}^{1} \int \psi_{k}^{\dag \,0} \widehat{V} \psi_{m}^{0}dpdq\nonumber\\
 &+&\sum_{k\neq n}a_{k}^{1}E_{n}^{1} +
  E_{n}^{2}\delta_{kn}\,. \notag\\
\end{eqnarray}

As before there are two possible cases, the first condition  $k=n$ implies
\begin{equation}
  E_{n}^{2}= \sum_{m\neq n}a_{m}^{1} \int \psi_{n}^{\dag \,0} \widehat{V} \psi_{m}^{0}dpdq
  -a_{n}^{1}E_{n}^{1} \notag
\end{equation}
If we set $a_{n}^{1}=0$ in the previous equation, then we have
\begin{eqnarray}
    E_{n}^{2}&=& \sum_{m\neq n}a_{m}^{1} \int \psi_{n}^{\dag \,0} \widehat{V} \psi_{m}^{0}dpdq \notag\\
    &=& \sum_{m\neq n}\frac{\int \psi_{m}^{\dag \,0} \widehat{V} \psi_{n}^{0}}{ E_{n}^{0} - E_{m}^{0} }dpdq
    \int \psi_{n}^{\dag \,0} \widehat{V} \psi_{m}^{0}dpdq \notag\\
    &=& \sum_{m\neq n}\frac{|\int \psi_{m}^{\dag \,0} \widehat{V} \psi_{n}^{0}|^{2}}{ E_{n}^{0} - E_{m}^{0} }dpdq\,.
     \label{eq:1.9}
\end{eqnarray}
The equation (\ref{eq:1.9}) is the second  order correction
to the energy of the unperturbed system.

\begin{figure}[htb!]
\includegraphics[scale=0.3]{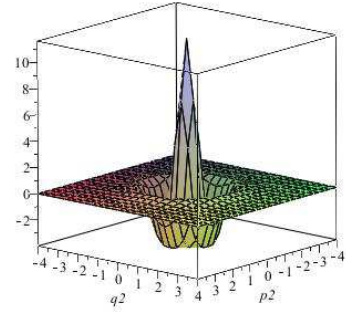}
\caption{Winger function second order n=0}
\label{fig12}
\end{figure}

\begin{figure}[htb!]
\includegraphics[scale=0.3]{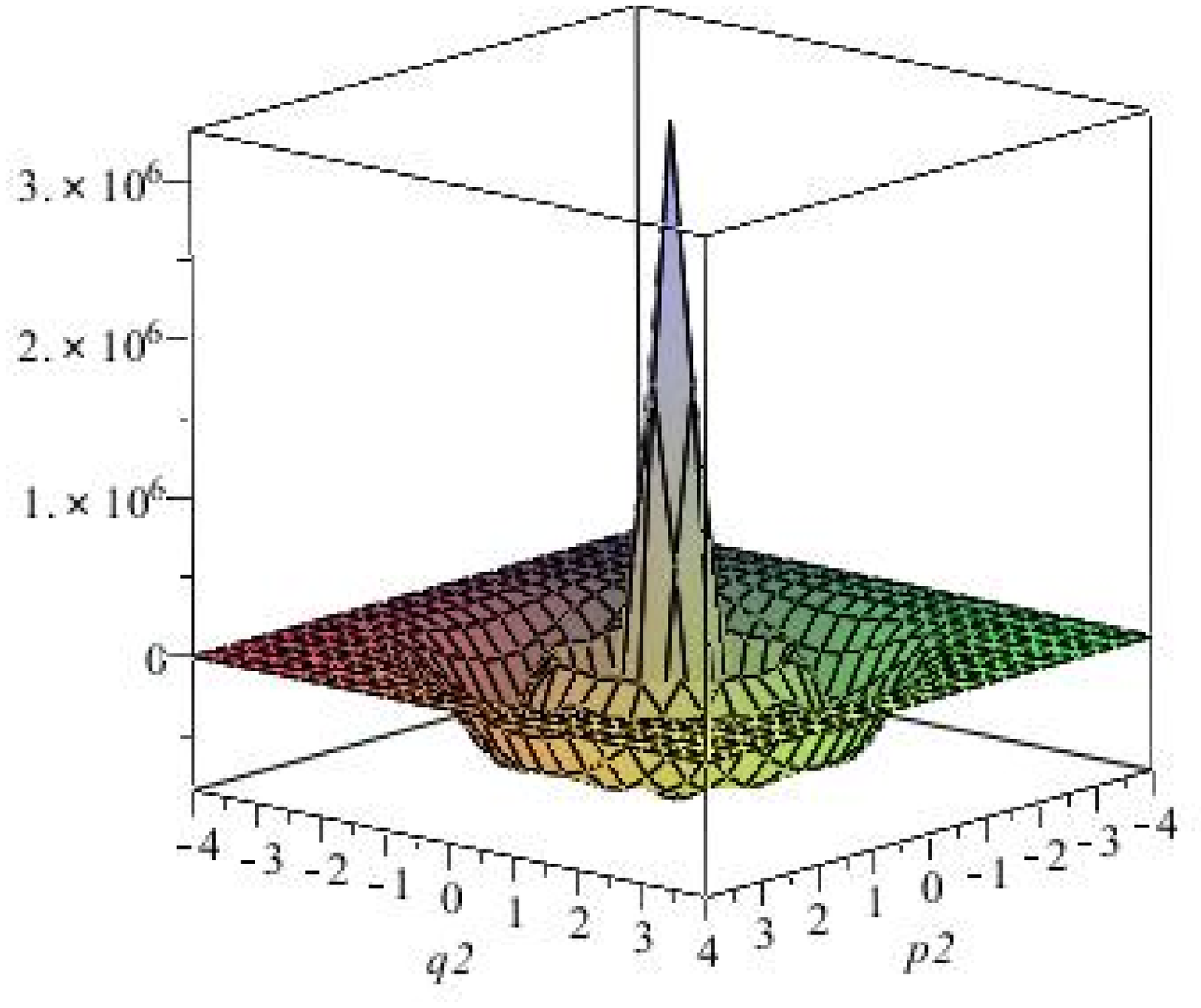}
\caption{Winger function second order n=2}
\label{fig22}
\end{figure}

\begin{figure}[htb!]
\includegraphics[scale=0.3]{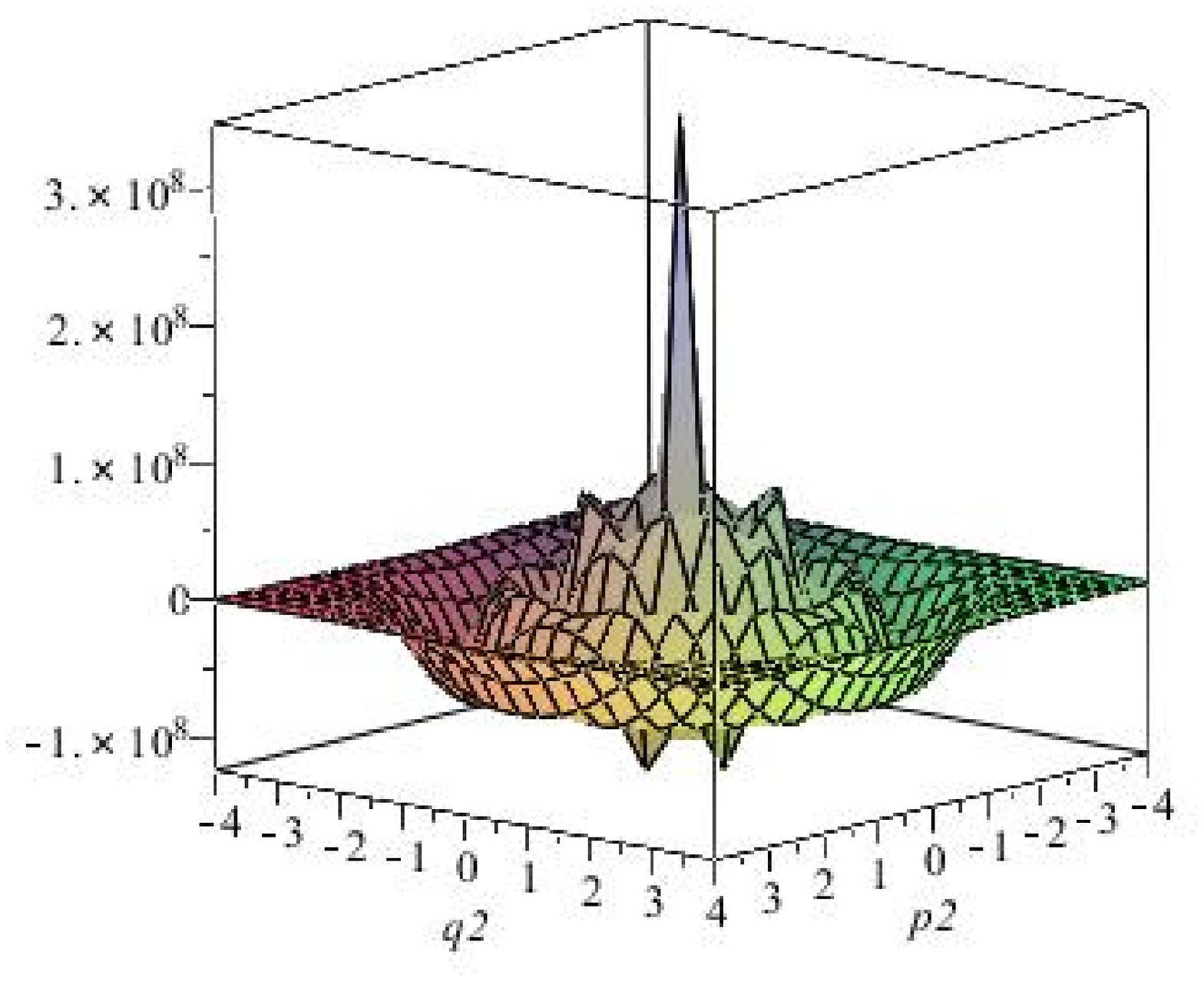}
\caption{Winger function second order n=4}
\label{fig32}
\end{figure}

\begin{figure}[htb!]
\includegraphics[scale=0.3]{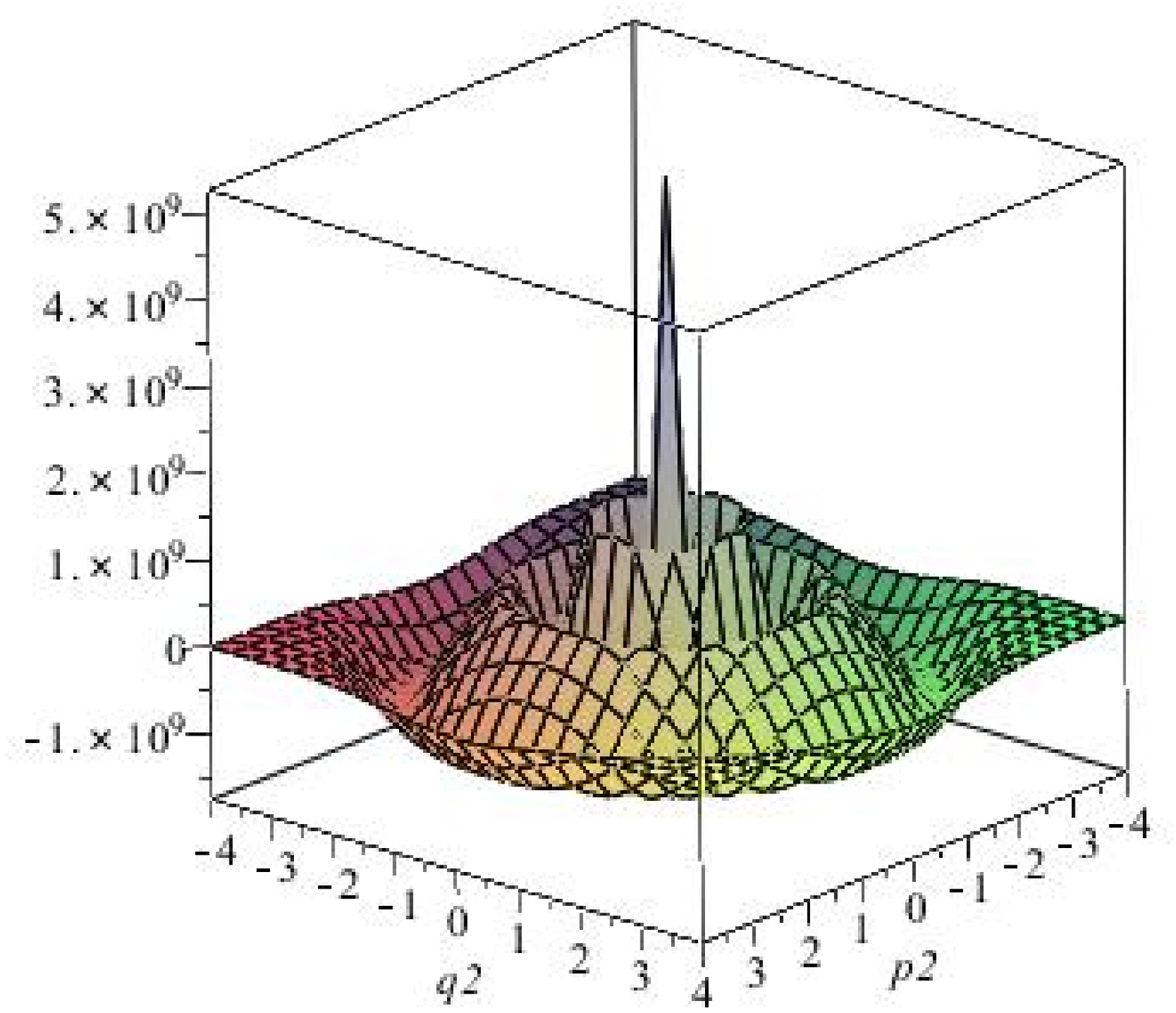}
\caption{Winger function second order n=6}
\label{fig42}
\end{figure}

\begin{figure}[htb!]
\includegraphics[scale=0.3]{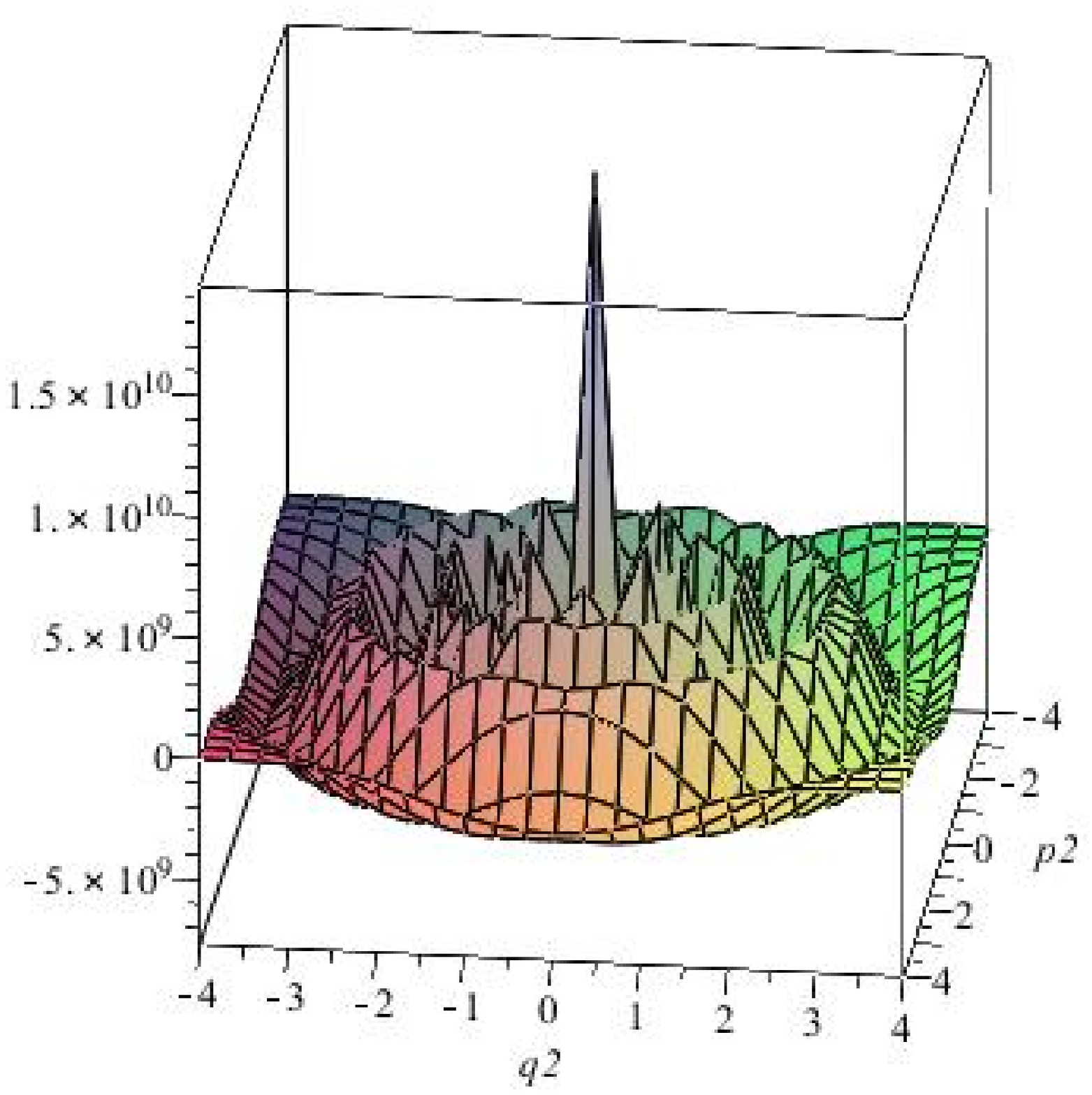}
\caption{Winger function second order n=8}
\label{fig52}
\end{figure}

The second condition $k\neq n$ leads to

\begin{eqnarray}
     a_{m}^{2}
    &=& \sum_{k\neq n} \int\frac{ \psi_{m}^{\dag \,0} \widehat{V} \psi_{k}^{0}}{(E_{n}^{0}-E_{m}^{0})}dpdq
    \int\frac{ \psi_{k}^{\dag \,0} \widehat{V} \psi_{n}^{0}}{ E_{n}^{0} - E_{k}^{0} }dpdq\notag \\
     &-&
    \int\frac{ \psi_{m}^{\dag \,0} \widehat{V} \psi_{n}^{0}}{(E_{n}^{0}-E_{m}^{0})^2}dpdq
    \int \psi_{n}^{\dag \,0} \widehat{V} \psi_{n}^{0}\,,  dpdq \notag\\
\end{eqnarray}
therefore
\begin{small}
\begin{eqnarray}
\psi_{n}^{2}&=& \sum_{m\neq n}a_{m}^{2}\psi_{m}^{0} \notag \\
 &=& \sum_{m\neq n}
 \bigg(\sum_{k\neq n} \int\frac{ \psi_{m}^{\dag \,0} \widehat{V} \psi_{k}^{0}}{(E_{n}^{0}-E_{m}^{0})}dpdq
    \int\frac{ \psi_{k}^{\dag \,0} \widehat{V} \psi_{n}^{0}}{ E_{n}^{0} - E_{k}^{0} }dpdq\notag \\
     &-&
    \int\frac{ \psi_{m}^{\dag \,0} \widehat{V} \psi_{n}^{0}}{(E_{n}^{0}-E_{m}^{0})^2}dpdq
    \int \psi_{n}^{\dag \,0} \widehat{V} \psi_{n}^{0}  dpdq \bigg)\psi_{m}^{0} \notag
\end{eqnarray}
\end{small}
As a consequence the second order Wigner function is given by
\begin{small}
\begin{equation*}
f_{n}^{2}(q_{x},p_{x},q_{y},p_{y})=\psi _{n_{x}}(q_{x},p_{x})\psi
_{n_{y}}(q_{y},p_{y})\ast \psi _{n_{x}}^{\dagger }(q_{x},p_{x})\psi
_{n_{y}}^{\dagger }(q_{y},p_{y})\,,
\end{equation*}
\end{small}
the results for some cases are presented in figures FIG.\ref{fig12}, FIG.\ref{fig22}, FIG.\ref{fig32}, FIG.\ref{fig42} and FIG.\ref{fig52}. We also chart some results on tables
TABLE \ref{tbl:tablelabel12}, TABLE \ref{tbl:tablelabel22}, TABLE \ref{tbl:tablelabel32}, TABLE \ref{tbl:tablelabel42}
 and TABLE \ref{tbl:tablelabel52}.


\begin{table}
\caption{Winger function second order n=0.}
\centering
\begin{tabular}{lll}
\hline
$\varepsilon$ & Maximum & Minimum  \\
\hline
0 & 6.165 &	-2.555 \\
0.28 & 6.518 &	-2.599\\
0.5 & 7.334 &-2.708\\
1 & 11.63 &	-3.995\\
\hline
\end{tabular}
\label{tbl:tablelabel12}
\end{table}

\begin{table}
\caption{Winger function second order n=2.}
\centering
\begin{tabular}{lll}
\hline
$\varepsilon$ & Maximum & Minimum  \\
\hline
0 & $3.214131\times10^6$ &	-847282.769\\
0.28 & $3.221852\times10^6$ &	-847109.82\\
0.5 & $ 3.238755\times10^6$ &	-846731.218\\
1 & $ 3.312671\times10^6$ &	-845075.61\\
\hline
\end{tabular}
\label{tbl:tablelabel22}
\end{table}

\begin{table}
\caption{Winger function second order n=4.}
\centering
\begin{tabular}{lll}
\hline
$\varepsilon$ & Maximum & Minimum  \\
\hline
0 & $ 3.41392\times10^8 $ & $-1.23009\times10^8 $ \\
0.28 & $3.419019\times10^8$ & $-1.231347\times10^8$ \\
0.5 & $ 3.43018\times10^8 $	&  $-1.234099\times10^8$ \\
1 & $3.47896\times10^8$ & $-1.246125\times10^8$\\
\hline
\end{tabular}
\label{tbl:tablelabel32}
\end{table}

\begin{table}
\caption{Winger function second order n=6.}
\centering
\begin{tabular}{lll}
\hline
$\varepsilon$ & Maximum & Minimum  \\
\hline
0 & $ 5.177845\times10^9$ &$-1.736374\times10^9$ \\
0.28 & $5.18489\times10^9$ &$-1.738067\times10^9$ \\
0.5 &$ 5.200312\times10^9$ 	&$-1.741771\times10^9$  \\
1 & $ 5.267715\times10^10$ 	&$-1.757962\times10^9$    \\
\hline
\end{tabular}
\label{tbl:tablelabel42}
\end{table}

\begin{table}
\caption{Winger function second order n=8.}
\centering
\begin{tabular}{lll}
\hline
$\varepsilon$ & Maximum & Minimum  \\
\hline
0 & $ 1.910526\times10^10$ &$	-7.538693\times10^9$ \\
0.28 &$ 1.913466\times10^10$	 &$	-7.549376\times10^9$ \\
0.5 &$ 1.919899\times10^10$ &$	-7.572756\times10^9$ \\
1 &$ 1.948017\times10^10$ &$	-7.674945\times10^9$\\
\hline
\end{tabular}
\label{tbl:tablelabel52}
\end{table}

\section{The non-classicality indicator}\label{nc}
The volume of negative part of Wigner function can be interpreted as a signature of quantum interference. In this way, a measure of non-classicality of quantum states is defined by negativity indicator, which is given by \cite{nc}
\begin{eqnarray}\label{d12}
\eta(\psi)&=&\int\int \left[|f_W\,_{\psi}(q,p)|-F_W\,_{\psi}(q,p)\right]dpdq\\ \nonumber
&=&\int\int |F_W\,_{\psi}(q,p)|dpdq -1.\nonumber
\end{eqnarray}
This indicator represents the doubled volume of the integrated part of the Wigner function. In sequence, we calculated numerically this indicator for one dimensional hydrogen atom. The results of this calculation are shown in Table \ref{tbl:tablelabelnc1} below. A interesting result is that the parameter $\eta(\psi)$ for $n=1$, the fundamental level of hydrogen atom, is equals to zero. This point shows that in this level the hydrogen atom can be studied by classical mechanics methods.

\begin{table}
\caption{One-dimensional Hydrogen atom.}
\centering
\begin{tabular}{ll}
\hline
$n$ & $\eta(\psi)$\\
\hline
1  & 0\\
 2 & 0.416375\\
3 & 0.693456\\
4 & 0.956789\\
5 & 1.217456\\
6 & 1.334907\\
7 & 1.491342\\
8 & 1.778512\\
9 & 1.901238\\
\hline
\end{tabular}
\label{tbl:tablelabelnc1}
\end{table}

In addition, we calculate numerically the non-classicality indicator for hydrogen atom with Henon-Heiles interaction. The results are shown in Tables TABLE \ref{tbl:tablelabelnc2}, TABLE \ref{tbl:tablelabelnc3}, TABLE \ref{tbl:tablelabelnc4} and TABLE \ref{tbl:tablelabelnc5}. An interesting result is the fact that the negativity indicator grows in the same direction of interaction parameter $\epsilon$. This latter result implies that the quantum entanglement of quantum system analyzed increases when the indicator parameter grows. It is important for quantum computing \cite{qc1}, for example.
\begin{table}
\caption{Negativity parameter for $\epsilon=0$.}
\centering
\begin{tabular}{ll}
\hline
$n$ & $\eta(\psi)$ \\
\hline
0  & 0\\
2 & 0.32645\\
4 & 0.45786\\
6 & 0.53647\\
8 & 0.60185\\
\hline
\end{tabular}
\label{tbl:tablelabelnc2}
\end{table}

\begin{table}
\caption{Negativity parameter for $\epsilon=0.28$.}
\centering
\begin{tabular}{ll}
\hline
$n$ & $\eta(\psi)$ \\
\hline
0  & 0.16783\\
2 & 0.35784\\
4 & 0.46210\\
6 & 0.54678\\
8 & 0.63193\\
\hline
\end{tabular}
\label{tbl:tablelabelnc3}
\end{table}

\begin{table}
\caption{Negativity parameter for $\epsilon=0.5$.}
\centering
\begin{tabular}{ll}
\hline
$n$ & $\eta(\psi)$ \\
\hline
0  & 0.19773\\
2 & 0.38954\\
4 & 0.47841\\
6 & 0.56823\\
8 & 0.67918\\
\hline
\end{tabular}
\label{tbl:tablelabelnc4}
\end{table}

\begin{table}
\caption{Negativity parameter for $\epsilon=1$.}
\centering
\begin{tabular}{ll}
\hline
$n$ & $\eta(\psi)$ \\
\hline
0  & 0.20376\\
2 & 0.39932\\
4 & 0.50385\\
6 & 0.58762\\
8 & 0.72431\\
\hline
\end{tabular}
\label{tbl:tablelabelnc5}
\end{table}

\section{Concluding remarks}\label{6}

The present work deals with the problem of constructing a formalism in phase
space based on the unitary symmetry. We have used the algebraic structure of
the Wigner function, along with the notion of star-products, to construct
unitary representations for the Galilei group. Then the physical aspects of
the formalism are discussed and, in particular, the physical meaning of a
quantum state is obtained by introducing the notion of quasi-amplitudes of
probability in $\Gamma $, which is related to the Wigner quasi-probability.
The Schr\"{o}dinger equation is constructed in phase space and as an application,
the Wigner function for the Hydrogen
is derived. Then we add a H\'enon-Heiles potential to the Hamiltonian of the Hydrogen atom and calculate corrections of Wigner function
at first and second order of pertubation. We presented our results in the panels and tables along the article.

\section*{Acknowledgements}

This work was partially supported by CNPq of Brazil and NSERC of Canada. Jos\'e Cruz thanks Department of Physics and Astronomy for access to various facilities
that made it easy for me to carry out research during one year stay in Victoria.

\appendix\label{appendix}
\section{Calculation of $a_n$}
In this appendix we present the coefficients $a_n$.

\begin{widetext}
\begin{eqnarray}
a0&=& \sqrt{n_{x}(n_{x}-1)}\psi_{n_{x}-2}^{(0)} \sqrt{n_{y}}%
\psi_{n_{y}-1}^{(0)}-\sqrt{n_{x}(n_{x}-1)}\psi_{n_{x}-2}^{(0)}  \sqrt{n_{y}+1}\psi_{n_{y}+1}^{(0)}\notag \\
&-&\sqrt{(n_{x}+1)(n_{x}+2)}%
\psi_{n_{x}+2}^{(0)} \sqrt{n_{y}}\psi_{n_{y}-1}^{(0)}
+\sqrt{(n_{x}+1)(n_{x}+2)}\psi_{n_{x}+2}^{(0)} \sqrt{n_{y}+1}%
\psi_{n_{y}+1}^{(0)} \,,\notag
\end{eqnarray}
\begin{eqnarray}
a1&=&  \frac{1}{3} \sqrt{n_{y}(n_{y}-1)(n_{y}-2)%
}\psi_{n_{y}-3}^{(0)}+ \bigg(\sqrt{n_{y}(n_{y}+1)^{2}} + \sqrt{n_{y}^{3}}+\sqrt{n_{y}(n_{y}-1)^{2}}\bigg)\psi_{n_{y}-1}^{(0)} \notag \\
&-& \bigg(\sqrt{(n_{y}+1)(n_{y}+2)^{2}} + \sqrt{(n_{y}+1)^{3}}+\sqrt{n_{y}^{2}(n_{y}+1)}\bigg)\psi_{n_{y}+1}^{(0)}
\notag \\
&-&\frac{1}{3} \sqrt{(n_{y}+1)(n_{y}+2)(n_{y}+3)}\psi_{n_{y}+3}^{(0)}\,,\notag
\end{eqnarray}
\begin{eqnarray}
a2&=& \bigg(\frac{1}{8} \sqrt{n_{x}(n_{x}-1)(n_{x}-2)(n_{x}-3)%
}\psi_{n_{x}-4}^{(0)}
+\frac{1}{4} \sqrt{n_{x}(n_{x}+1)^2(n_{x}-1)%
}\psi_{n_{x}-2}^{(0)} \notag \\
&+&\frac{1}{4} \sqrt{n_{x}^{3}(n_{x}+1) }\psi_{n_{x}-2}^{(0)}\bigg)
\sqrt{n_{y}(n_{y}-1)%
}\psi_{n_{y}-2}^{(0)}\notag \\
&+& \sqrt{n_{x}(n_{x}-1)%
}\psi_{n_{x}-2}^{(0)}\bigg(\frac{1}{8} \sqrt{n_{y}(n_{y}-1)(n_{y}-2)(n_{y}-3)%
}\psi_{n_{y}-4}^{(0)}\notag \\
&+&\frac{1}{4} \sqrt{n_{y}(n_{y}+1)^2(n_{y}-1)%
}\psi_{n_{y}-2}^{(0)}+\frac{1}{4} \sqrt{n_{y}^{3}(n_{y}-1) }\psi_{n_{y}-2}^{(0)}\bigg)\,,\notag
\end{eqnarray}
\begin{eqnarray}
a3&=& \sqrt{n_{x}(n_{x}-1)%
}\psi_{n_{x}-2}^{(0)}\bigg(\frac{1}{4} \sqrt{n_{y}(n_{y}-1)^{3}%
}\psi_{n_{y}-2}^{(0)}-\frac{1}{4} \sqrt{(n_{y}+1)(n_{y}+2)(n_{y}+3)^2%
}\psi_{n_{y}+2}^{(0)}\notag \\
&+&\frac{1}{4} \sqrt{n_{y}(n_{y}-1)(n_{y}-2)^2%
}\psi_{n_{y}-2}^{(0)}-\frac{1}{4} \sqrt{(n_{y}+1)(n_{y}+2)^3 }\psi_{n_{y}-2}^{(0)}\bigg)\notag\\
&+&\bigg(-\frac{1}{8} \sqrt{n_{x}(n_{x}-1)(n_{x}-2)(n_{x}-3)%
}\psi_{n_{x}-4}^{(0)}-\frac{1}{4} \sqrt{n_{x}(n_{x}-1)(n_{x}+2)^2%
}\psi_{n_{x}-2}^{(0)} \notag \\
&-&\frac{1}{4} \sqrt{n_{x}^{3}(n_{x}-1) }\psi_{n_{x}-2}^{(0)}\bigg)
\sqrt{(n_{y}+1) +(n_{y}+2)%
}\psi_{n_{y}+2}^{(0)}\notag \\
&+&\sqrt{n_{x}(n_{x}-1)%
}\psi_{n_{x}-2}^{(0)}\bigg(-\frac{1}{4} \sqrt{(n_{y}+1)^{3}(n_{y}+2)%
}\psi_{n_{y}+2}^{(0)}-\frac{1}{4} \sqrt{n_{y}^{2}(n_{y}+1)(n_{y}+2)%
}\psi_{n_{y}+2}^{(0)}\notag \\
&-&\frac{1}{8} \sqrt{(n_{y}+1)(n_{y}+2)(n_{y}+3)(n_{y}+4)%
}\psi_{n_{y}+4}^{(0)}\bigg)\,,\notag
\end{eqnarray}
\begin{eqnarray}
a4&=&
\bigg(\frac{1}{4} \sqrt{n_{x}(n_{x}-1)^3%
}\psi_{n_{x}-2}^{(0)}-\frac{1}{4} \sqrt{(n_{x}+1)(n_{x}+2)(n_{x}+3)^2%
}\psi_{n_{x}+2}^{(0)}\bigg) \sqrt{n_{y}(n_{y}-1)}\psi_{n_{y}-2}^{(0)}   \notag \\
&+&\bigg(-\frac{1}{4} \sqrt{n_{x}(n_{x}-1)^{3} }\psi_{n_{x}-2}^{(0)} +
\frac{1}{4} \sqrt{(n_{x}+1)(n_{x}+2)(n_{x}+3)^2%
}\psi_{n_{x}+2}^{(0)} \bigg)
\sqrt{(n_{y}+1)(n_{y}+2)%
}\psi_{n_{y}+2}^{(0)}\notag \\
%
&+&\bigg(\frac{1}{4} \sqrt{n_{x}(n_{x}-1)(n_{x}-2)^2%
}\psi_{n_{x}-2}^{(0)}-\frac{1}{4} \sqrt{(n_{x}+1)(n_{x}+2)^3%
}\psi_{n_{x}+2}^{(0)}\bigg) \sqrt{n_{y}(n_{y}-1)}\psi_{n_{x}-2}^{(0)}   \notag \\
&+&\bigg(-\frac{1}{4} \sqrt{n_{x}(n_{x}-1)(n_{x}-2)^{2} }\psi_{n_{x}-2}^{(0)} +
\frac{1}{4} \sqrt{(n_{x}+1)(n_{x}+2)^3%
}\psi_{n_{x}+2}^{(0)} \bigg)
\sqrt{(n_{y}+1)(n_{y}+2)%
}\psi_{n_{y}+2}^{(0)}\notag \\
&+&\bigg(-\frac{1}{4} \sqrt{(n_{x}+1)^3(n_{x}+2)%
}\psi_{n_{x}+2}^{(0)}-\frac{1}{4} \sqrt{n_{x}^{2}(n_{x}+1)(n_{x}+2)%
}\psi_{n_{x}+2}^{(0)} \notag \\
 &-&\frac{1}{8} \sqrt{(n_{x}+1)(n_{x}+2)(n_{x}+3)(n_{x}+4)%
}\psi_{n_{x}+4}^{(0)} \bigg) \sqrt{n_{y}(n_{y}-1)}\psi_{n_{x}-2}^{(0)} \,,  \notag
\end{eqnarray}
\begin{eqnarray}
a5&=&
 \sqrt{(n_{x}+1)(n_{x}+2)%
}\psi_{n_{x}+2}^{(0)}\bigg(-\frac{1}{8} \sqrt{n_{y}(n_{y}-1)(n_{y}-2)(n_{y}-3)%
}\psi_{n_{y}-4}^{(0)}\notag \\
&-&\frac{1}{4} \sqrt{n_{y}(n_{y}-1)(n_{y}+1)^2%
}\psi_{n_{y}-2}^{(0)}
-\frac{1}{4} \sqrt{n_{y}^3(n_{y}-1)%
}\psi_{n_{y}-2}^{(0)}
-\frac{1}{4} \sqrt{n_{y}(n_{y}-1)^3%
}\psi_{n_{y}-2}^{(0)} \notag \\
 &+&\frac{1}{4} \sqrt{(n_{y}+1)(n_{y}+2)(n_{y}+3)^2%
}\psi_{n_{y}+2}^{(0)} -\frac{1}{4} \sqrt{n_{y}(n_{y}-1)(n_{y}-2)^2%
}\psi_{n_{y}-2}^{(0)} \notag \\
 &+&\frac{1}{4} \sqrt{(n_{y}+1)(n_{y}+2)^3%
}\psi_{n_{y}+2}^{(0)}
\bigg) \,,\notag
\end{eqnarray}
\begin{eqnarray}
a6&=&
\bigg(\frac{1}{4} \sqrt{(n_{x}+1)^3(n_{x}+2)%
}\psi_{n_{x}+2}^{(0)} + \frac{1}{4} \sqrt{n_{x}^2(n_{x}+1)(n_{x}+2)%
}\psi_{n_{x}+2}^{(0)} \notag \\
&+&\frac{1}{8} \sqrt{(n_{x}+1)(n_{x}+2)(n_{x}+3)(n_{x}+4)%
}\psi_{n_{x}+4}^{(0)}\bigg) \sqrt{(n_{y}+1)(n_{y}+2)}\psi_{n_{y}+2}^{(0)}   \notag \\
&+& \sqrt{(n_{x}+1)(n_{x}+2)%
}\psi_{n_{x}+2}^{(0)}\bigg(\frac{1}{4} \sqrt{(n_{y}+1)^3(n_{y}+2)%
}\psi_{n_{y}+2}^{(0)}\notag \\
&+&\frac{1}{4} \sqrt{n_{y}^2(n_{y}+1)(n_{y}+2)%
}\psi_{n_{y}+2}^{(0)}
+\frac{1}{8} \sqrt{(n_{y}+1)(n_{y}+2)(n_{y}+3)(n_{y}+4)%
}\psi_{n_{y}+4}^{(0)
}\bigg) \notag
\end{eqnarray}
and
\begin{eqnarray}
a7&=&
\bigg(\frac{1}{2} \sqrt{n_{x}(n_{x}-1)%
}\psi_{n_{x}-2}^{(0)} -\frac{1}{2} \sqrt{(n_{x}+1)(n_{x}+2)}\psi_{n_{x}+2}^{(0)}%
+\frac{1}{2} \sqrt{n_{y}(n_{y}-1)%
}\psi_{n_{y}-2}^{(0)}\notag\\
&-&\frac{1}{2} \sqrt{(n_{y}+1)(n_{y}+2)%
}\psi_{n_{y}+2}^{(0)}\bigg) \,.\notag
\end{eqnarray}
\end{widetext}

\end{document}